\documentclass[aps,pra,showpacs,showkeys,floatfix,twocolumn,byrevtex,superscriptaddress]{revtex4-1}

\usepackage{amsmath}
\usepackage{amssymb}
\usepackage{amstext}
\usepackage{amsopn}
\usepackage{amsfonts}
\usepackage{amsxtra}
\usepackage{hyperref} 
\usepackage{cleveref}
\usepackage[english]{babel}
\usepackage{multirow}
\usepackage{color}
\usepackage{verbatim}
\usepackage{graphicx}
\usepackage{dcolumn}
\usepackage{bm}

\graphicspath{{./}{figs//}}
\newcounter{saveenumi}

\newcommand{\be}{\begin{enumerate}}
	\newcommand{\ee}{\end{enumerate}}

\newcommand*{\rom}[1]{\expandafter\@slowromancap\romannumeral #1@}

\begin{document}

\title{Scaling Behaviour of Low-Temperature Orthorhombic Domains in Prototypical High-Temperature Superconductor La$_{1.875}$Ba$_{0.125}$CuO$_{4}$}

\author{T.~A.~Assefa}\email[E-mail me at: ]{tassefa@bnl.gov}
\affiliation{Condensed Matter Physics and Materials Science Department, Brookhaven National Laboratory, Upton, New York~11973, USA}
\author{Y.~Cao}
\affiliation{Condensed Matter Physics and Materials Science Department, Brookhaven National Laboratory, Upton, New York~11973, USA}
\affiliation{Materials Science Division, Argonne National Laboratory, Lemont, IL 60439, USA}
\author{J.~Diao}
\affiliation{London Center for Nanotechnology, University College London, London~WC1E~6BT, UK}
\author{K.~Kisslinger}
\affiliation{Center for Functional Nanomaterials, Brookhaven National Laboratory, Upton, NY~11793, USA}
\author{G.~D.~Gu}
\affiliation{Condensed Matter Physics and Materials Science Department, Brookhaven National Laboratory, Upton, New York~11973, USA}
\author{J.~M.~Tranquada}
\affiliation{Condensed Matter Physics and Materials Science Department, Brookhaven National Laboratory, Upton, New York~11973, USA}
\author{M.~P.~M.~Dean}
\affiliation{Condensed Matter Physics and Materials Science Department, Brookhaven National Laboratory, Upton, New York~11973, USA}
\author{I.~K.~Robinson}\email[E-mail me at: ]{irobinson@bnl.gov}
\affiliation{Condensed Matter Physics and Materials Science Department, Brookhaven National Laboratory, Upton, New York~11973, USA}
\affiliation{London Center for Nanotechnology, University College London, London~WC1E~6BT, UK}

\date{\today}


\begin{abstract}
Translational/rotational symmetry breaking and recovery in condensed matter systems are closely related to exotic physical properties such as superconductivity (SC), magnetism, spin density waves (SDW) and charge density waves (CDW). The interplay between different order parameters is intricate and often subject to intense debate, as in the case of CDW order and superconductivity. In La$_{1.875}$Ba$_{0.125}$CuO$_{4}$~(LBCO), the locations of CDW domains are found to be pinned on the nanometer size scale. Coherent X-ray diffraction techniques open routes to directly visualize the domain structures associated with these symmetry changes. We have pushed Bragg Coherent Diffractive Imaging (BCDI) into the cryogenic regime where most phase transitions in quantum materials reside. Utilizing BCDI, we image the structural evolution of LBCO microcrystal samples during the high-temperature-tetragonal (HTT) to low-temperature-orthorhombic (LTO) phase transition. Our results show the formation of LTO domains close to the transition temperature and how the domain size varies with temperature. The LTO domain size is shown to decrease with temperature and to be inversely proportional to the magnitude of the orthorhombic distortion. The number of domains follows the secondary order parameter (or orthorhombic strain) measurement with a critical exponent that is consistent with the 3D universality class. 
\end{abstract}


\maketitle


\section{\label{sec:intro}Introduction}

Transition metal oxides (TMOs) host interesting structural and electronic transitions as well as the emergent phenomenon of high-temperature superconductivity\cite{Kivelson2003,Dagotto2001,Keimer2015,Tranquada2008}. In addition, these materials host charge, spin and orbital orders which are coupled to the lattice and are believed to compete with superconductivity in interesting ways\cite{Hucker2011}. Many of these TMOs exhibit charge density wave (CDW) phases, characterized by periodic modulation of conduction electrons, which suppress/co-exist with superconductivity\cite{Comin2016,Tranquada1995,Li2007}. 

The prototypical example of a high-temperature superconductors with competing CDW order is La$_{1.875}$Ba$_{0.125}$CuO$_{4}$~(LBCO). The occurrence of CDW order is closely tied to the lattice symmetry. LBCO has a high temperature tetragonal (HTT) crystal structure at room temperature, switching upon cooling to a low-temperature-orthorhombic~(LTO) structure with a transition temperature of about 240~K, then followed by a low-temperature-tetragonal~(LTT) phase with a transition temperature of 54~K. The HTT phase is characterized by untilted CuO$_{6}$ octahedra; in LTO they are tilted along the Cu-Cu bond direction, and in the LTT phase they are tilted along the Cu-O bond direction \cite{Hucker2011,Bozin2015,Axe1989,Achkar2016}. Like in other second-order phase transitions\cite{Goldstone1962}, the HTT-LTO structural transition is mediated by a soft phonon mode\cite{Birgeneau1987}. This structural phase transition has been studied with different X-ray/neutron scattering \cite{Hucker2011,Bozin2015,Fabbris2013} and transmission electron microscopy techniques\cite{Zhu1994,Horibe1997,Horibe2000}. These structural phase transitions can be modeled by solving the Ginzburg-Landau (GL) free-energy functional by considering a symmetry-invariant combination of order parameters\cite{Axe1989,Landau1937}. Derived physical properties such as the orthorhombic strain tensor and the coupling with a local order parameter can be extracted after minimizing the GL free-energy\cite{Ting1993}.  

Long range CDW order occurs only in the LTT phase of LBCO\cite{Fujita2004,Hucker2011,Wilkins2011,Miao2017,Miao2019}. Soft X-ray coherent scattering experiments have shown that the CDW phase is quite static\cite{Chen2016}, and the CDW pinning landscape is inherited from a domain wall structure of the LTO phase\cite{Chen2019}. Recently, a speckle correlation analysis on (012)$_{\textrm{LTO}}$ diffraction peak showed, the diffraction patterns changed whenever the sample heated above the HTT-LTO transition temperature, indicating the LTO domains are not spatially pinned in space\cite{Robinson2019}. Characterizing the LTO domains in three dimensions, is therefore relevant in understanding the physics behind the pinning phenomenon. In this regard, Bragg coherent diffraction imaging (BCDI) is an ideal probe technique to look at samples of micron-size scale, giving a real-space image of the domain texture in these materials. In a far-field limit, the diffraction pattern collected from a finite crystal is related to the Fourier transform of the electron density. It is difficult to recover the image of electron density since we lost the phase information. However, in BCDI technique the inverse phase problem can be retrieved given a diffraction pattern is oversampled with respect to the spatial Nyquist-Shannon sampling frequency\cite{Shannon1949} and with suitable phase retrieval algorithms\cite{Fienup1978,Gerchberg1972}. When a nanocrystal is illuminated with coherent X-rays with a coherence volume larger than the sample, then the diffraction pattern contains interference from all its regions appearing as fringes surrounding the Bragg peak in three dimensions\cite{Miao1999,Robinson2001,Vartanyants2001}. The simplest fringes are those which arise from interference between opposing facets on the crystal shape. Then diffraction patterns are collected with small rotations of the crystal, which comprises a complete three-dimension map of reciprocal space around the Bragg reflection. Then with appropriate choice of iterative algorithms the complete three-dimensional data can be inverted to give the electron density of the crystal\cite{Robinson2009}

In this paper, we present 3D renderings of LTO domains within an LBCO single crystal sample, obtained using the BCDI technique\cite{Robinson2009,Robinson2016,Pfeifer2006}. Inverted images show the formation of domains as we cool down the sample below the HTT-LTO phase transition temperature. A slice through the rendered inverted images shows the internal structure of LTO domains and the domain walls formed along the [110]$_{\textrm{HTT}}$ as a stack. In addition, from reconstructed images, we estimate the LTO domain size to be between 200-400~nm at 228~K, which is consistent with TEM results\cite{Horibe2000}.

\section{\label{sec:Experiments}Experimental methods}
\begin{figure}[ht!]
	\includegraphics[trim={2.cm 0 0 0},clip,scale=0.7]{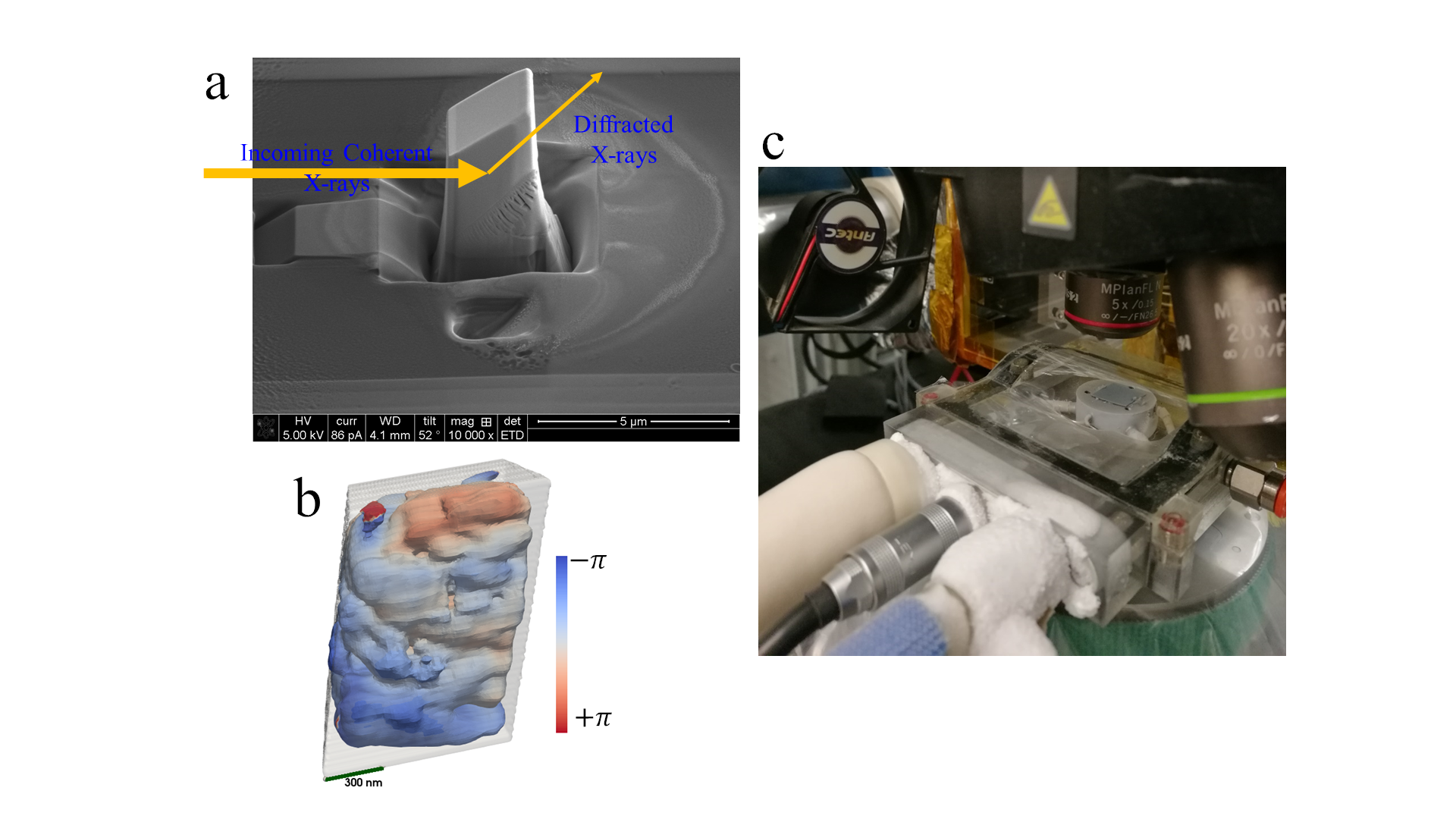}
	\caption{\label{fig:Fig1}(a) Scanning Electron Microscopy (SEM) image of the 1.6x1.6x1.6$\mu$m$^{3}$ cube-shaped LBCO sample which was cut-out of a pre-oriented single crystal using the FIB milling process. The arrows indicate the direction of the incoming and diffracted X-ray beam. (b) Isosurface rendering of the reconstructed image obtained after phase retrieval.(C) Picture of the Linkam stage during the BCDI experiment.}
\end{figure}

A high-quality single crystal of LBCO was prepared by the floating zone method\cite{Gu1993}. To obtain the micron size crystal needed for the BCDI study, the large LBCO crystal was oriented crystallographically using a Laue diffractometer. Then a 1.6$\times$1.6$\times$1.6$\mu$m$^{3}$ cube sample was cut out from the pre-oriented crystal via the in-situ lift-out method utilizing the Omniprobe manipulation system and Field Electron and Ion (FEI) Helios 600 dual-beam focused-ion-beam (FIB) (See Fig.~\ref{fig:Fig1})\cite{Hofmann2017}. The size of the cube was chosen to be less than the extinction depth of 9 keV X-rays in LBCO to minimize dynamical diffraction effects\cite{Hu2018,Civita2018,Shabalin2017}. Then the sample was welded with Pt onto a silicon wafer. This procedure was carried out at the Center for Functional Nanomaterials (CFN) at Brookhaven National Laboratory (BNL). In addition to the Pt-wielding, a solution of 2$~\%$ Tetraethyl-orthosilicate (TEOS) in ethanol was drop-casted on the LBCO cube and then annealed for about 5 hours at about 700 K in an oxygen atmosphere to avoid loss of oxygen during the annealing process. This method has been used for small metal nanocrystals, and has proved an important step to keep the nanocrystals fixed during transportation and measurements\cite{Monteforte2016}. Then the sample was mounted on a custom modified Linkam stage for BCDI measurements where the flow of liquid nitrogen can be controlled precisely by the T96 controller to set a specific target temperature and program linear cooling ramps up to  373 K/min. The complete cooling system has a controller, a pump, the Linkam stage, and a liquid nitrogen holder Dewar. The system allows cooling down to about 173 K, and low cooling rates give less icing and reduce noise and vibration.

Bragg coherent diffraction data were collected at the 34-ID-C beamline of the Advanced Photon Source (APS). For the coherent diffraction experiments, a monochromatic and Coherent X-ray beam size of 30(vertical)$\times$40(horizontal) $\mu$m$^{2}$ illuminated the LBCO cube sample. The X-ray beam size was chosen mainly because the 1.6 $\mu$m$^{3}$ cube sample is larger than the focused X-ray beam at 34-ID-C. Since the sample was pre-aligned, the precise crystal alignment was quickly determined. Then Coherent X-ray Diffraction (CXD) patterns from the (103)$_{\textrm{HTT}}$ and (114)$_{\textrm{HTT}}$ Bragg peaks were acquired using a Timepix detector mounted at 2m away from the sample. The full sensor of the detector has 512$\times$512 pixels with a pixel size of 55 $\mu$m. Diffraction data were collected at each step while rocking the sample in increments of 0.0025$^{\circ}$ around the Bragg peak. Before feeding the CXD data to an iterative phasing algorithm developed in Matlab\cite{Robinson2009,Pfeifer2006,Robinson2001,Robinson2008}, both whitefield correction and hot pixel removal were applied for each diffraction pattern. For the phasing, a combination of error-reduction (ER) and Hybrid-input-output (HIO) algorithms\cite{Gerchberg1972,Fienup1982} have been used alternately, with the iteration starting and ending with ER. The well-defined shapes/edges of the sample help to render the diffraction patterns invertible, which also allows us to use a fixed box-shaped support to assist the phasing algorithms. This is an essential experimental advancement because the soft edges of even the best-focused X-ray beams are currently thought to be insufficiently sharp to use as support constraints\cite{Fienup1978,Fienup1987}. Moreover, when the particle size is larger than both the longitudinal and transverse coherence length, the reconstructed images tend to have artifacts such as non-uniform amplitude distribution, with fewer facets and missing parts\cite{Leake2009}. In our case, this was mitigated by turning on the Partial Coherence Correction (PCC) in the iterative phasing algorithm at iteration ten and then turned off at about one-third way through the total iteration numbers\cite{Clark2012,Vartanyants2001,Search2001}.

\section{\label{sec:BCDIresults}Bragg Coherent Diffraction Imaging Results}

\begin{figure}[ht!]
	\includegraphics[trim={1.8cm 1cm 0.5cm 0.8cm},clip,scale=0.7]{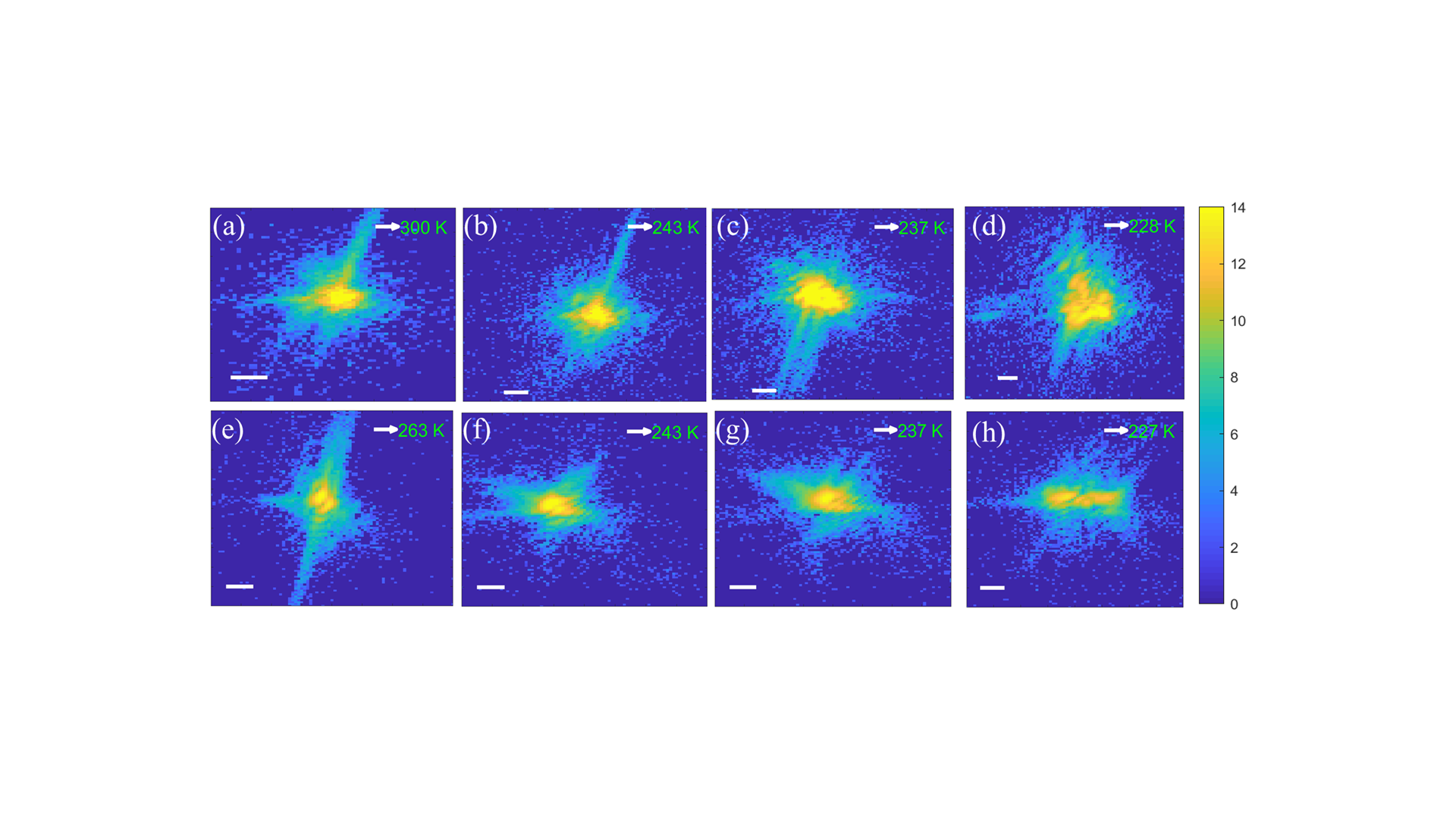}
	\caption{\label{fig:Fig2}(a)-(f) Logarithmic-scale plots of (a)-(d) the (103)$_{\textrm{HTT}}$ Bragg peak and (e)-(f)  the (114)$_{\textrm{HTT}}$ Bragg peak of the LBCO sample measured at different temperatures. The scale bar shown is 40 pixels corresponding to 1$\times$10$^{-1}$nm$^{-1}$. Both the $\delta$q$_{x}$ and $\delta$q$_{y}$ are mutually perpendicular reciprocal space vectors coplanar to the CCD surface and calculated as (2$\pi$/$\lambda$)(p/D), where $\lambda$=1.3776$\AA$ is the wavelength, p=55 $\mu$m is detector pixel size and D=2 m is sample-to-detector distance.}
\end{figure}

Our CXD results from the (103)$_{\textrm{HTT}}$ and (114)$_{\textrm{HTT}}$ structural Bragg peaks show similar behavior. As shown in Fig.~\ref{fig:Fig2}, both diffraction peaks are split on the detector, which is an indication of LTO twin-domain formation as reported from previous X-ray \cite{Axe1989} and electron diffraction measurements\cite{Zhu1994,Horibe1997,Chen1991,Chen1993,Chen1999}. Plots of the coherent diffraction patterns collected near the (103)$_{\textrm{HTT}}$ and (114)$_{\textrm{HTT}}$ Bragg reflections from the same sample are shown in Fig.~\ref{fig:Fig2}(a-d) and (e-h) respectively. Initially, both (103)$_{\textrm{HTT}}$ and (114)$_{\textrm{HTT}}$ diffraction peaks broaden as the sample temperature decreases. Then both the (103)$_{\textrm{HTT}}$ and (114)$_{\textrm{HTT}}$ diffraction peaks become split when the temperature falls below 240(5)~K. At all temperatures, both diffraction peaks are strongly speckled on the detector because of the high coherence of the beam and stability of the 34-ID-C setup. The diffraction peak splitting temperature is consistent with the HTT-LTO transition temperature, reported from previous X-ray and neutron measurements\cite{Hucker2011,Bozin2015,Axe1989}.

\begin{figure}[h!]
	\includegraphics[trim={4cm 1.3cm 0.6cm 0.6cm},clip,scale=1]{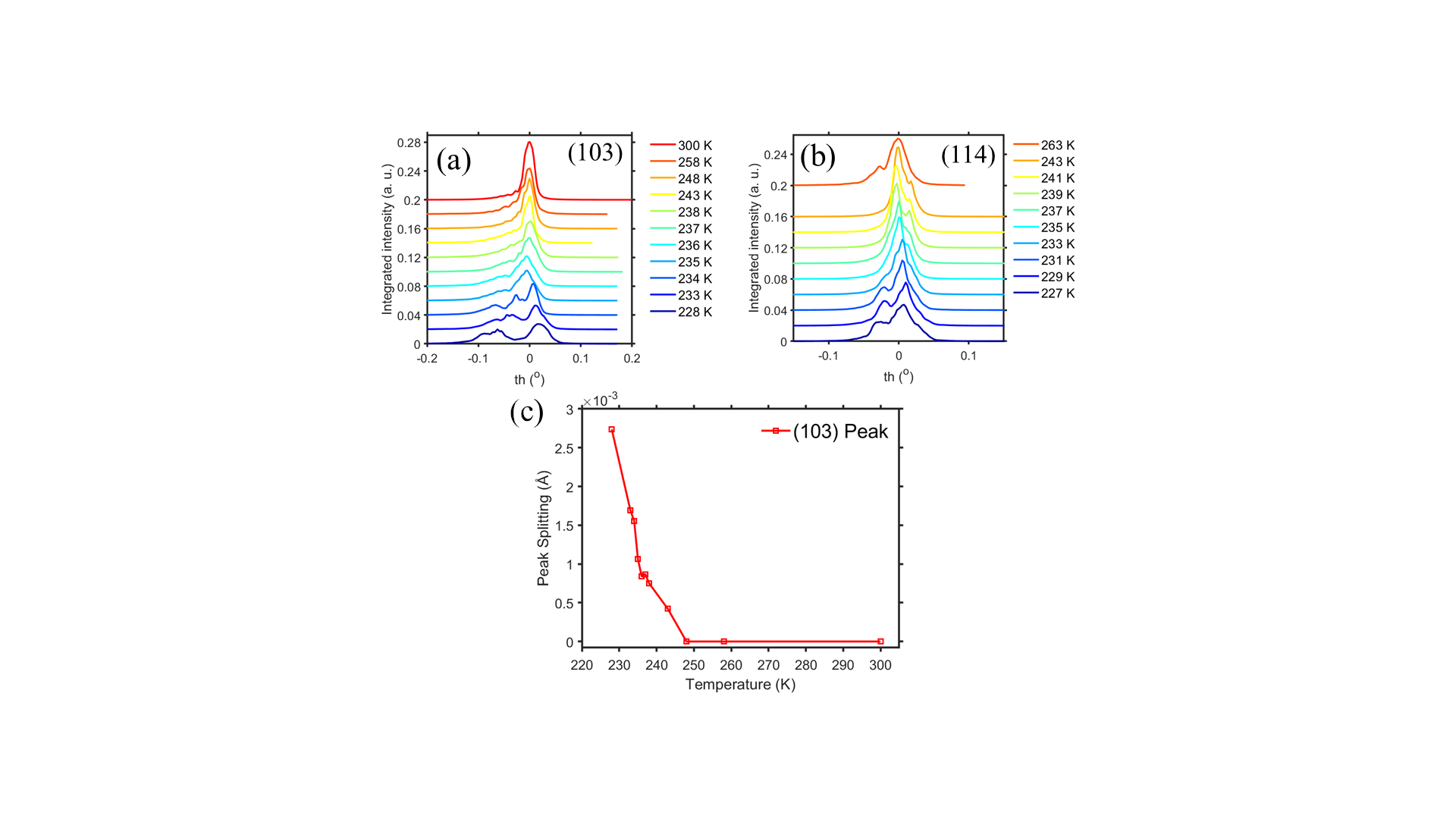}
	\caption{\label{fig:Fig3}(a) and (b) rocking curves of (103)$_{\textrm{HTT}}$ and (114)$_{\textrm{HTT}}$ integrated, background subtracted Bragg peaks as a function of temperature crossing the HTT to LTO phase transition. (c) Total (103) diffraction peak splitting in three dimensions.}
\end{figure}

Moreover, three-dimensional (3D) diffraction data were collected from both Bragg reflections at several temperatures spanning the HTT to LTO phase transition. From the white-field and flat-field corrected images two regions of interest, ROI1 and ROI2 were integrated over the Bragg peak and far away for the background subtraction, respectively, and difference plotted as rocking curves, shown in Fig.~\ref{fig:Fig3}(a) and (b). Similar to what we observed in the 2D diffraction data, the rocking curves also show peak splitting. The (103)$_{\textrm{HTT}}$ peak split into multiple peaks on the detector as the sample temperature decreases continuously, whereas the (114)$_{\textrm{HTT}}$ peak has a tiny peak in the left side the rocking curve in the HTT phase, indicating inhomogeneity possibly introduced during ion-milling of the sample. As a result, both the peak splitting analysis and reconstruction will focus on the (103)$_{\textrm{HTT}}$ Bragg peak data. To calculate the total peak splitting displacement $\Delta q$ for the (103)$_{\textrm{HTT}}$ peak; first, we recorded the difference in pixel position $\Delta p_{\textrm{x}}$ and $\Delta p_{\textrm{y}}$ on the detector and the frame number $\Delta p_{\textrm{z}}$ for all temperatures. We convert the difference in pixels and frame number to $\textrm{\AA}^{-1}$ as 
$\Delta q_x = (2 \pi / \lambda)(p/D) \Delta p_x$,
$\Delta q_y = (2 \pi / \lambda)(p/D) \Delta p_y$, and
$\Delta q_z = Q \Delta \theta \Delta p_z$,
where $\lambda$ is the X-ray wavelength, $Q$ is the momentum transfer, $p$ is the pixel size, $D$ is the detector distance and $\Delta \theta$ is the step size of the rocking scan.
Finally, the three dimensional peak splitting shown in Fig.~\ref{fig:Fig3}(c) is calculated as $\Delta$q = $\sqrt{(\Delta q_{x}^2 + \Delta q_{y}^2 + \Delta q_{z}^2)}$. Figure~\ref{fig:Fig3}(c) shows that the peak splitting disappears at the expected HTT-LTO transition temperature indicating the formation of (113)$_{\textrm{LTO}}$ rotated twins domains. Moreover, the splitting onset temperature determined from the rocking curves shown in Figs.~\ref{fig:Fig3}(a) \& (b) differs slightly between the (103)$_{\textrm{HTT}}$ peak at 234 K and (114)$_{\textrm{HTT}}$ at 235 K, which we attribute to a finite uncertainty in the measurement such as temperature off-set between the sample and temperature recorded by Linkam cooling stage sensor.


\begin{figure}[ht!]
	\includegraphics[trim={4.cm 0.95cm 1.5cm 0.9cm},clip,scale=0.9]{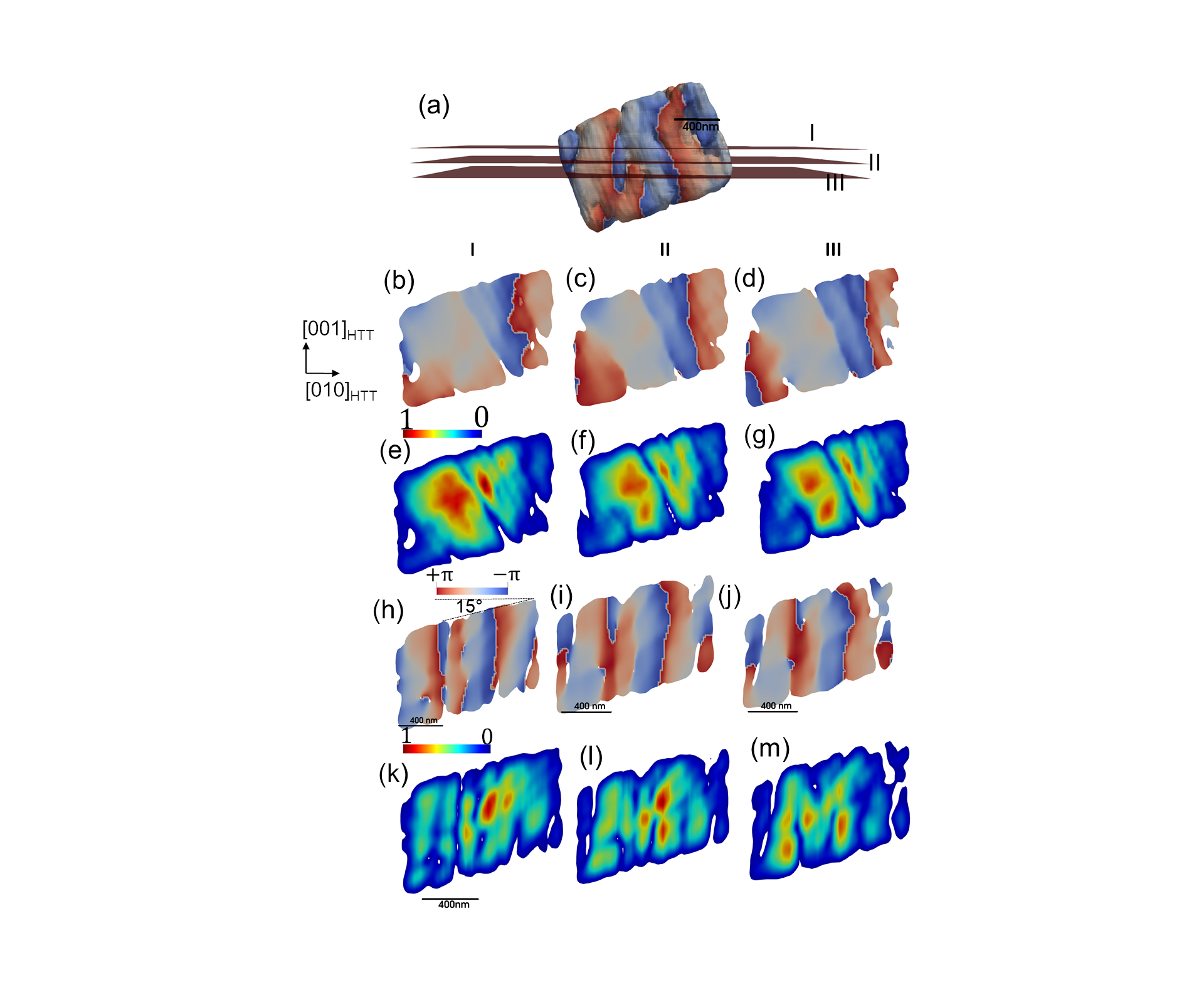}
	\caption{\label{fig:Fig4}  Three-dimensional (3D) isosurface rendering plotted at about 10\% of the small LBCO single-crystal reconstructed from three-dimensional diffraction patterns using the phase retrieval algorithm. The slice in [100] plane indicated in as I, II, and III. (b)-(d) and (e)-(g) are slices at different positions of the crystal showing a map of the image phase (projection of the lattice displacement) and amplitude (electron density) for HTT phase respectively. Similarly, (h)-(j) and (k)-(m) are slices at different positions of the crystal showing a map of the image phase (projection of the lattice displacement) and amplitude (electron density) for LTO phase respectively.}
\end{figure}

We interpret the peak splitting as due to \textit{a/b} twinning in the orthorhombic phase of LBCO and can use BCDI to obtain images of the pattern of domains in three dimensions. In order to visualize the evolution of LTO domain formation close to the transition temperature, we inverted the 3D coherent diffraction patterns using iterative phasing. The reconstruction results in Fig.~\ref{fig:Fig4} show a clear difference between the LTO and HTT phases reconstructed from the 228~K and 258~K temperature data. To understand better the internal structure of the phases, we take a slice cut through the reconstructed image in [100] plane. At 258~K the slice shows a ''single'' domain whereas at 228~K, it shows the presence of several domains with sizes in the range of 150-350~nm. There is a phase ramp between the domains which has a size of 20-50~nm. These domain and domain-wall sizes are close to those reported in electron microscopy studies\cite{Zhu1994,Horibe2000,Li2003,Chen1991}. To make a comparison with TEM dark-field results we take a slice along the [001] plane of 228 K data and the result is shown in Fig.\ref{fig:FigS2} of supplementary material. Similar layer like domains are observed, which are elongated in one direction. The domain and domain-wall sizes obtained from our reconstruction results are in the same order of magnitude with TEM results. Furthermore, we present the 235 K data in a similar fashion (see Fig.\ref{fig:FigS1} of supplementary material); the data shows the domain size is larger than the 228 K data and domains are stacked in a similar way. To look at the amplitude and phase variation inside the reconstructed images of HTT and LTO phases, a line-profiles through [001] slices are shown in \ref{fig:FigS3} and \ref{fig:FigS4}. 


Because the crystal is isolated at the center of the diffractometer, samples prepared through FIB gives us an opportunity of measuring multiple peaks from the same crystal, without any contaminating signals from neighboring crystals. This in the future has a potential application for quantum materials where one can image a single FIB crystal using both structural and electronic order peaks and overlay reconstructed real-space images. However, the ion milling process can also introduce undesired damage, amorphization layer, and strain on the surface of the sample, or can affect the chemical composition. Typically, the damage of the FIB'ed sample is 20 to 30~nm for 30~keV Ga ions, and 5~kev ions would have three times less effect\cite{Kato2004}. Also, how far the Ga ions penetrate the sample depends on both the energy of ions and the angle of polishing (normal incidence versus glancing incidence)\cite{Hofmann2018}. For gold nanocrystals, Ga ions can go up to 50~nm at 30~keV and normal incidence and decreases a factor of five at 5~keV and glancing incidence \cite{Hofmann2018}. Although we used 5~keV ion beam for final polishing of the present sample to minimize the damage, amorphization layer and strain, some of the strains and non-sharp edges could be partly due to the beam milling process.

\begin{figure}[ht!]
	\includegraphics[trim={0.5cm 0.5cm 0.5cm 0.5cm},clip,scale=0.35]{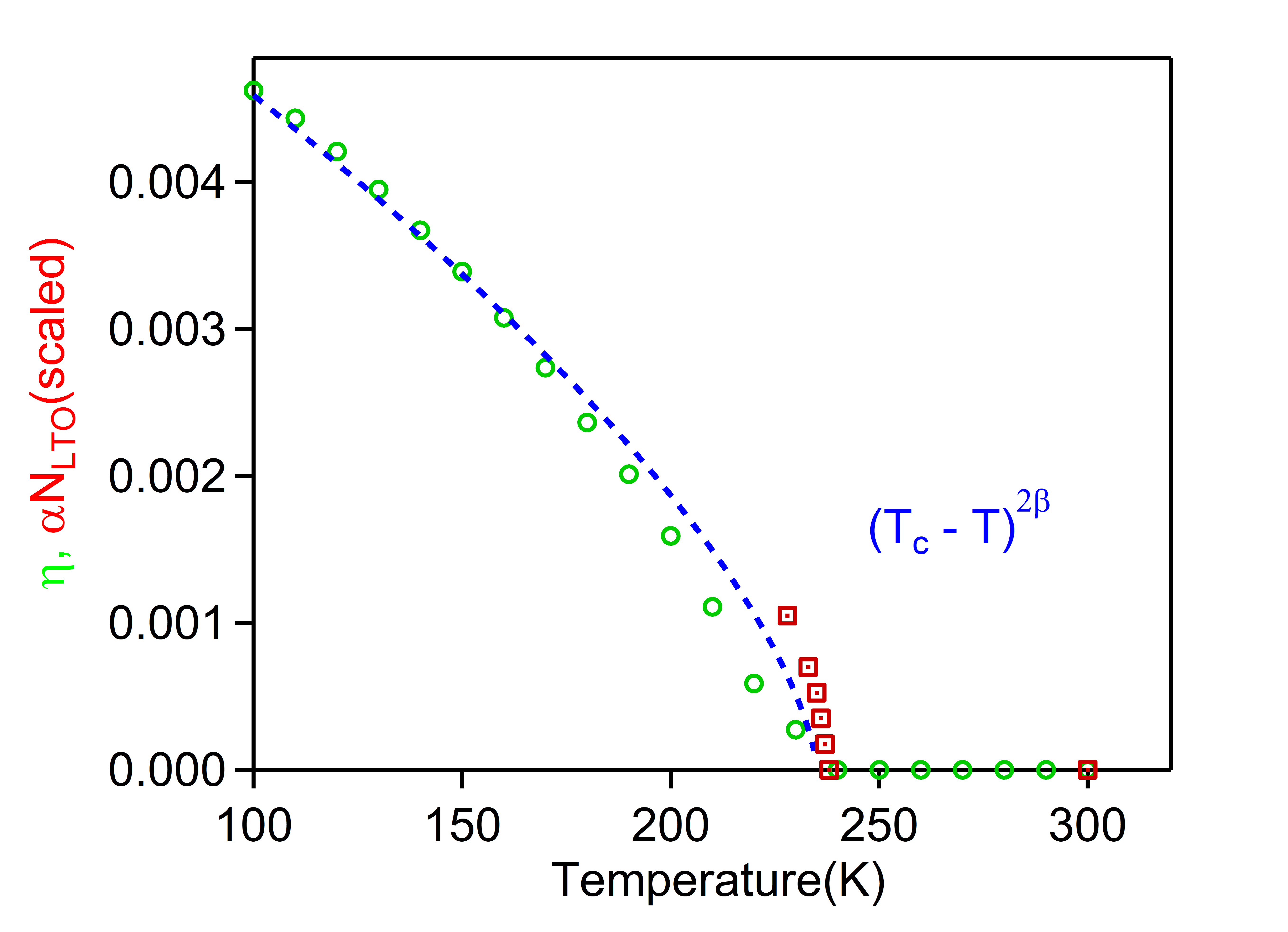}
	\caption{\label{fig:Fig5} Temperature evolution of lattice orthorhombicity strain ($\eta$) calculated as $\frac{2(a-b)}{(a+b)}$, the lattice parameters are derived from Rietveld refinements\cite{Bozin2015}(green circles). Temperature evolution of scaled number of domains is extracted from BCDI reconstruction and counting by visualizing in Paraview (red squares). We also plotted $\alpha$(T$_{c}$-T)$^{2\beta}$ function with T$_{c}$=238~K, $\alpha$=1.75$\times$10$^{-4}$ and $\beta$=0.33 as a blue dashed line which is guide to the eye.}
\end{figure}

BCDI reconstructed images allow us to count domains in three dimensions\cite{AHRENS2005}. Figure ~\ref{fig:Fig5} shows that the number of domains increases dramatically when the sample temperature is below the HTT-LTO transition temperature. An early Ginzburg-Landau (GL) study of the HTT-LTO transition has derived the critical behavior of the orthorhombicity (strain) near the LTO transition temperature with the critical exponent $\beta=$0.33\cite{Birgeneau1987}. The identical critical behavior of the number of domains, N$_{\textrm{LTO}}$, is observed here, indicating a linear relation between N$_{\textrm{LTO}}$ and $\phi^{2}$ \cite{Birgeneau1987}. This effect can be understood by adding the lowest symmetry allowed coupling term N$_{\textrm{LTO}}$ ($\Delta\phi^{2}$), that arises from the formation of domain walls, in the GL free energy\cite{Muller1971,Chaikin1995}.  


\subsection{\label{subsec:DomainIndexing}Domain Indexing by Fourier Filtering}

\begin{figure}[ht!]
	\includegraphics[trim={5.4cm 0.7cm 4.cm 0.6cm},clip,scale=1.6]{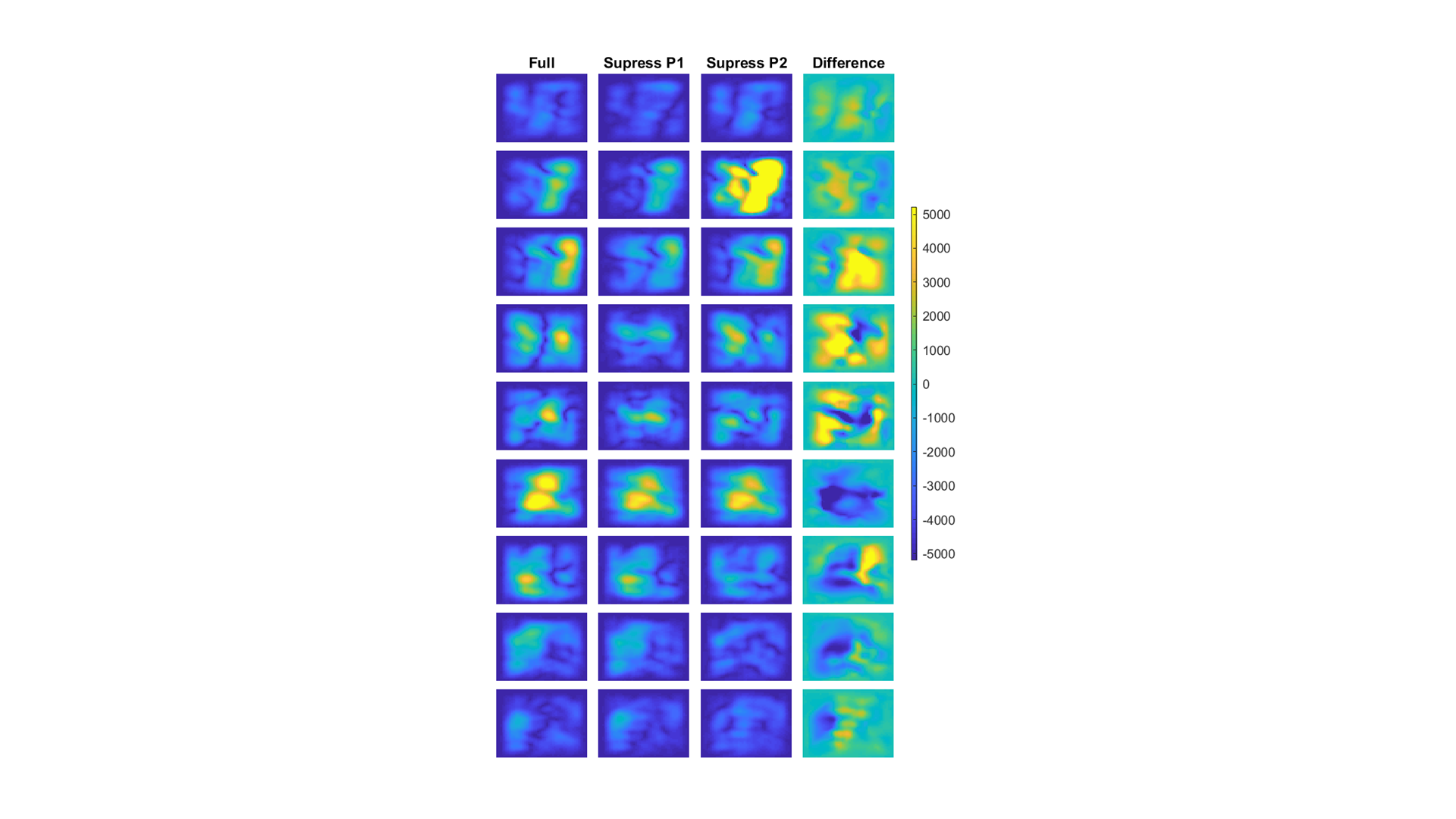}
	\caption{\label{fig:Fig6}Fourier filtering on the reconstructed image measured at 228~K of the different slices along the third dimension. Suppress P1 is obtained by Fourier filtering of $21\times21\times25$ voxels around the smaller peak by setting it to zero for reconstruction. Suppress P2 are obtained by Fourier filtering of $13\times13\times13$ voxels around the larger peak by setting to zero for reconstruction.}
\end{figure}

\begin{figure}[ht!]
	\includegraphics[trim={1.9cm 0.8cm 1cm 1.cm},clip,scale=0.7]{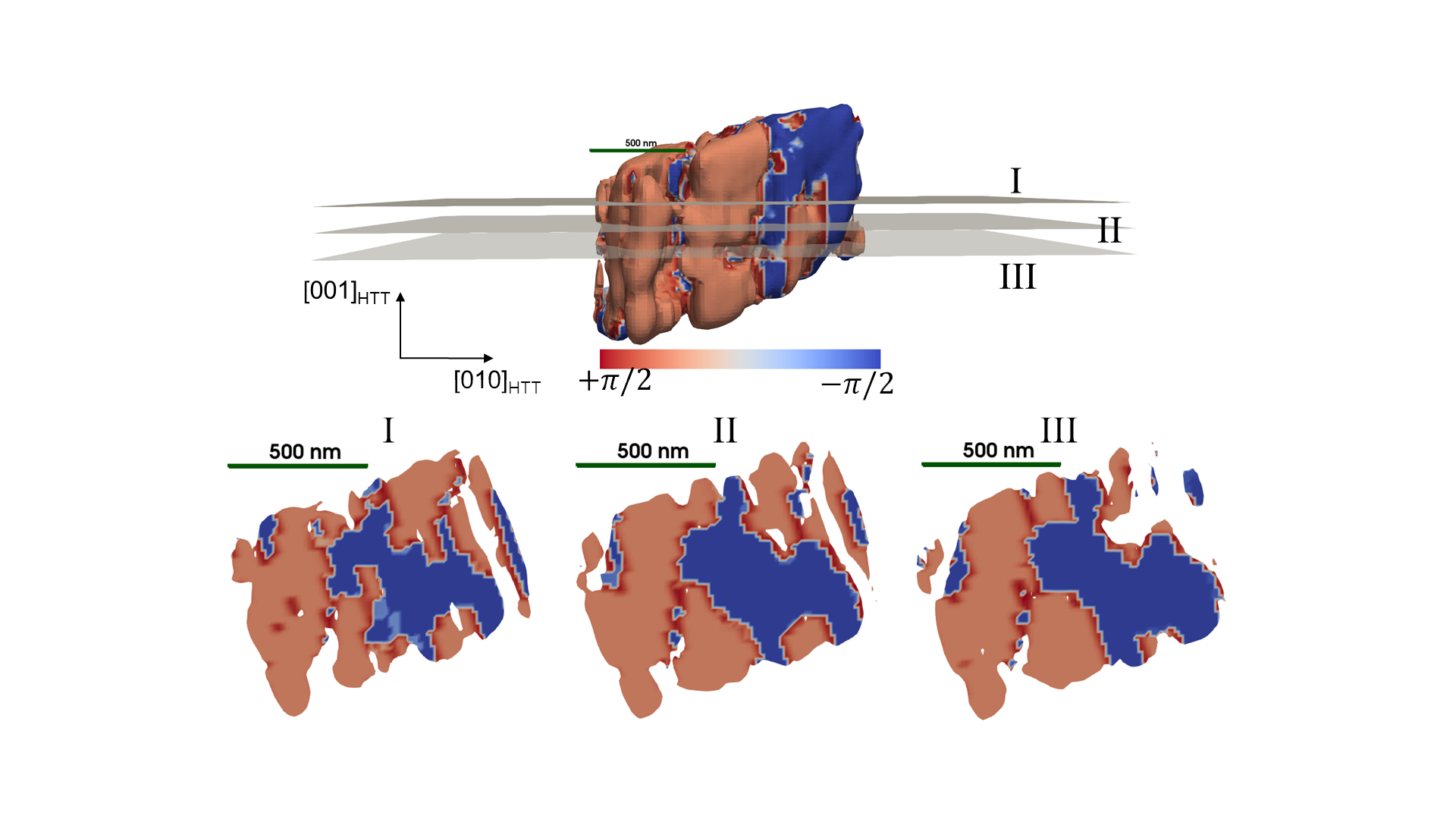}
	\caption{\label{fig:Fig7} Isosurface rendering plotted at 10\% of reconstructed images for the 228~K data shown after indexed domains. The I, II, and III are sliced across the reconstructed image to show the internal structure of domains.}
\end{figure}

The complex pattern of domains seen in the low-temperature orthorhombic phase is believed to be due to twinning between regions of opposite orientation of their $a$ and $b$ axes\cite{Bratkovsky1996}. The diffraction peak splitting at low temperatures arises for the same reason. With BCDI, we have the unique opportunity to assign which domain in the image arises from which peak in the diffraction pattern.  This is undertaken in Figs.~\ref{fig:Fig6} and \ref{fig:Fig7} to test the idea.

The final 3D image of the domains at the lowest temperature, measured at the $(103)_{\textrm{HTT}}$ diffraction peak was Fourier transformed back to reciprocal space to regenerate the split diffraction peak, but retaining all the phase information. A region of $21\times21\times25$ voxels was set to zero around the first diffraction peak and it was inverse Fourier transformed to give an image with all the domains contributing to that peak suppressed. This was repeated for the second Bragg peak by setting $13\times13\times13$ voxels to zero.The results are shown in Fig.~\ref{fig:Fig6} in the raw coordinate system of the discrete Fourier transform of the data voxels, ($x$, $y$) detector pixels and $z$ steps on the rocking curve. The (x,y,z) directions are roughly aligned with the Cartesian (x,y,z) coordinate system used in Fig 4.  The z-slices shown in Fig 6, show cross-sections of the sample roughly perpendicular to the X-ray beam direction. The image amplitude is presented on the same color scale of  0 to 1.6$\times10^{4}$ in the first three columns for a selection of $z$ slices whereas the difference between Suppress P1 and Suppress P2 is shown in a scale of -5.2$\times10^{4}$ to 5.2$\times10^{4}$. It can be seen that different parts of the initial domain image (left) become reduced in amplitude in the two derived images of Suppress P1 and Suppress P2.

To visualize the domain identities more clearly, a color image was generated in the same physical "laboratory" Cartesian coordinate frame already used to present the images in Fig.~\ref{fig:Fig4}. Here, $z$ runs along the beam, $x$ is horizontal, transverse to the beam, and $y$ is vertical. In Fig.~\ref{fig:Fig7} an isosurface af the crystal and three slices through the 3D image are shown corresponding to the views of Fig.~\ref{fig:Fig4} (k,l,m). The domains are colored red or blue according to whether their amplitude is higher with the first or second peak suppressed.

There is a clear pattern in these images where domains are color-coded according to the diffraction peak to which they contribute most.  It appears that one end of the crystal mostly contributes to the "blue" peak and the other end to the "red" peak. In between, there is some alternation of domain identities, as expected from the microtwinning concept \cite{Bratkovsky1996}. This result strongly supports the picture of twinning underlying to formation of domains in the LBCO tetragonal to orthorhombic phase transition.

\section{\label{sec:Conclusion}Conclusion and outlook}

Coherent X-ray diffraction technique allows us to track the evolution of structural domains by monitoring a shared Bragg diffraction pattern on a 2D detector. The transition temperature deduced agrees with previously published X-ray studies\cite{Bozin2015,Hucker2011}. Moreover, the speckle pattern in reciprocal space is a unique finger-print of how domains are staggered in the sample (real-space) and both $(103)_{\textrm{HTT}}$ and $(114)_{\textrm{HTT}}$ diffraction patterns split into multiple peaks indicating the formation of twin domains. This agrees with recent speckle correlation analysis results on (012)$_{\textrm{LTO}}$ diffraction peak\cite{Robinson2019}.     

In BCDI technique, the phase information lost during the measurement is retrieved with the computational technique with properly oversampled diffraction patterns; one can iteratively reconstruct the phase. For weak phasing objects such as metal nanoparticles\cite{Robinson2009,Pfeifer2006,Newton2010}, battery materials\cite{Singer2018,Ulvestad2015}, and oxides\cite{Robinson1999}, the retrieved real-space images give internal strain information in addition to electron density, which is not accessible with any other technique\cite{Miao2015}. However, imaging structural texture of strongly correlated materials presents a challenge to the technique and obtaining a unique solution is very challenging. To circumvent this issue, we implement a fixed-box support constraint in the iterative phasing algorithm which allowed us to invert the reciprocal diffraction patterns to real-space images and gave a reproducible result. The reconstructed real-space images of domains we observe are LTO twin domains which are very common for this type of sample. Neighboring domains show a phase shift, and the phase difference between the two nearby domains gives the relative displacement of twin domain walls. The observation of LTO domains agrees with previous "microstructures" (domains) of La$_{2-x}$Sr$_{x}$CuO$_{4}$\cite{Horibe1997, Horibe2000} and LTO La$_{2-x}$Ba$_{x}$CuO$_{4}$\cite{Zhu1994} obtained with dark field transmission electron microscopy. As shown in Fig.~\ref{fig:Fig5}, the number of domains follows a similar path as the degree of orthorhombicity (orthorhombic strain) derived from powder diffraction data\cite{Bozin2015}, which is related to the order parameter\cite{Chaikin1995}. This can be understood by adding a new domain-wall associated energy term in the GL free-energy equation. 

%
\begin{acknowledgments}
We want to thank Miao Hu, Daniel Mazzone, and Emil Bozin for their insight on the experimental results. We also would like to thank Ross Harder, Wonsuk Cha and Evan Maxey for their support during the beamtime and for their help setting up the Linkam cooling stage for experiments at 34-ID-C. TAA would like to thank Felix Hofmann for his comments on the experimental results and insights on focused ion-beam milling processes. X-ray experiments are supported by the U.S. Department of Energy, Office of Science, Basic Energy Sciences, Materials Science and Engineering Division, under Contract No. DE-SC0012704(BNL) and DE-AC02-06CH11357(ANL). The focused-ion beam sample preparation used the resources of Center for Functional Nanomaterials, which is a U.S. DOE Office of Science Facility, at BNL under Contract No. DE-SC0012704. The experiments were carried out at the Advanced Photon Source(APS) beamline 34-ID-C, and the APS was supported by the U. S. Department of Energy, Office of Science, Office of Basic Energy Sciences, under Contract No. DE-AC02-06CH11357. The beamline 34-ID-C was built with U.S. National Science Foundation grant DMR-9724294.  
\end{acknowledgments}

\bibliography{ImagingLTODomain}
\bibliographystyle{apsrev}
\bibliographystyle{jabbrv_abbrv}

\vfill\newpage
\renewcommand\thefigure{S\arabic{figure}}
\setcounter{figure}{0}

\pagebreak
\widetext
\begin{center}
	\textbf{\large Supplemental Materials: Scaling Behaviour of Low-Temperature Orthorhombic Domains in Prototypical High-Temperature Superconductor La$_{1.875}$Ba$_{0.125}$CuO$_{4}$}
\end{center}

\setcounter{equation}{0}
\setcounter{figure}{0}
\setcounter{table}{0}
\setcounter{page}{1}
\makeatletter
\renewcommand{\theequation}{S\arabic{equation}}
\renewcommand{\thefigure}{S\arabic{figure}}
\renewcommand{\bibnumfmt}[1]{[S#1]}
\renewcommand{\citenumfont}[1]{S#1}

\section{Supplementary Figures}
\label{sec:SM}


\begin{figure}[htb]
	\includegraphics[trim={2.5cm 1.5cm 1.5cm 1.8cm},clip,scale=1.6]{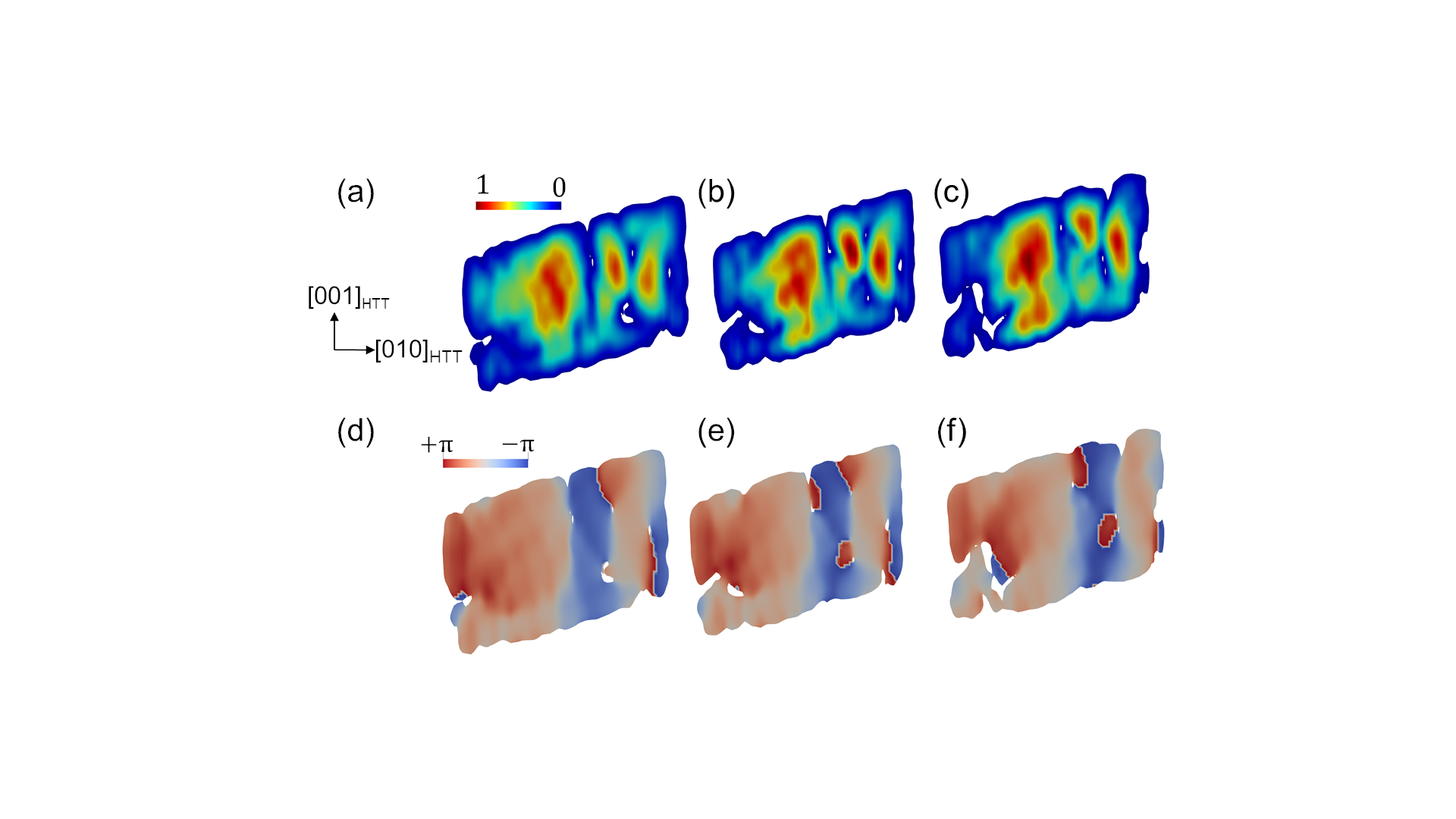}
	\caption{\label{fig:FigS1} Two-dimensional slices of [100] plane of the 235 K data. The slices are taken at the same positions, as shown in Fig.\ref{fig:Fig4}(a). (a-c) and (d)-(f) are sliced at different positions of the crystal showing a map of the amplitude (electron density) and phase (projection of the lattice displacement) respectively.}
\end{figure}

\vspace{2cm}
\begin{figure}[htb]
	\includegraphics[trim={2.5cm 1.5cm 1.5cm 1.8cm},clip,scale=1.6]{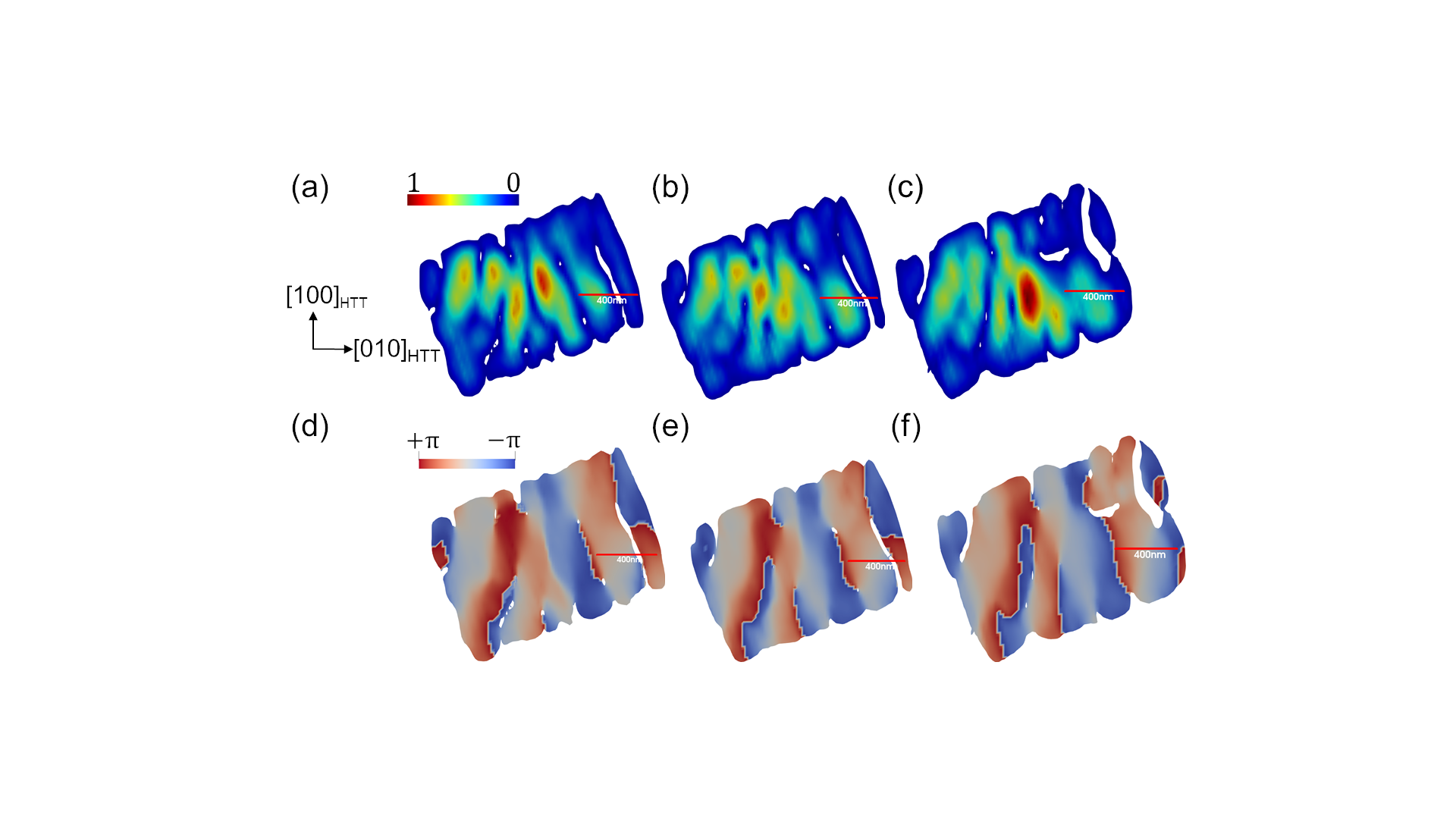}
	\caption{\label{fig:FigS2} Two-dimensional slices of the [001] plane of the 228 K image. The slices are taken at different positions in the crystal, similar to those shown in Fig.4(a). (a-c) and (d)-(f) are sliced at different positions of the crystal showing a map of the amplitude (electron density) and phase (projection of the lattice displacement) respectively.}
\end{figure}

\begin{figure}[h]
	\includegraphics[trim={2.5cm 1.5cm 1.5cm 1.5cm},clip,scale=1.6]{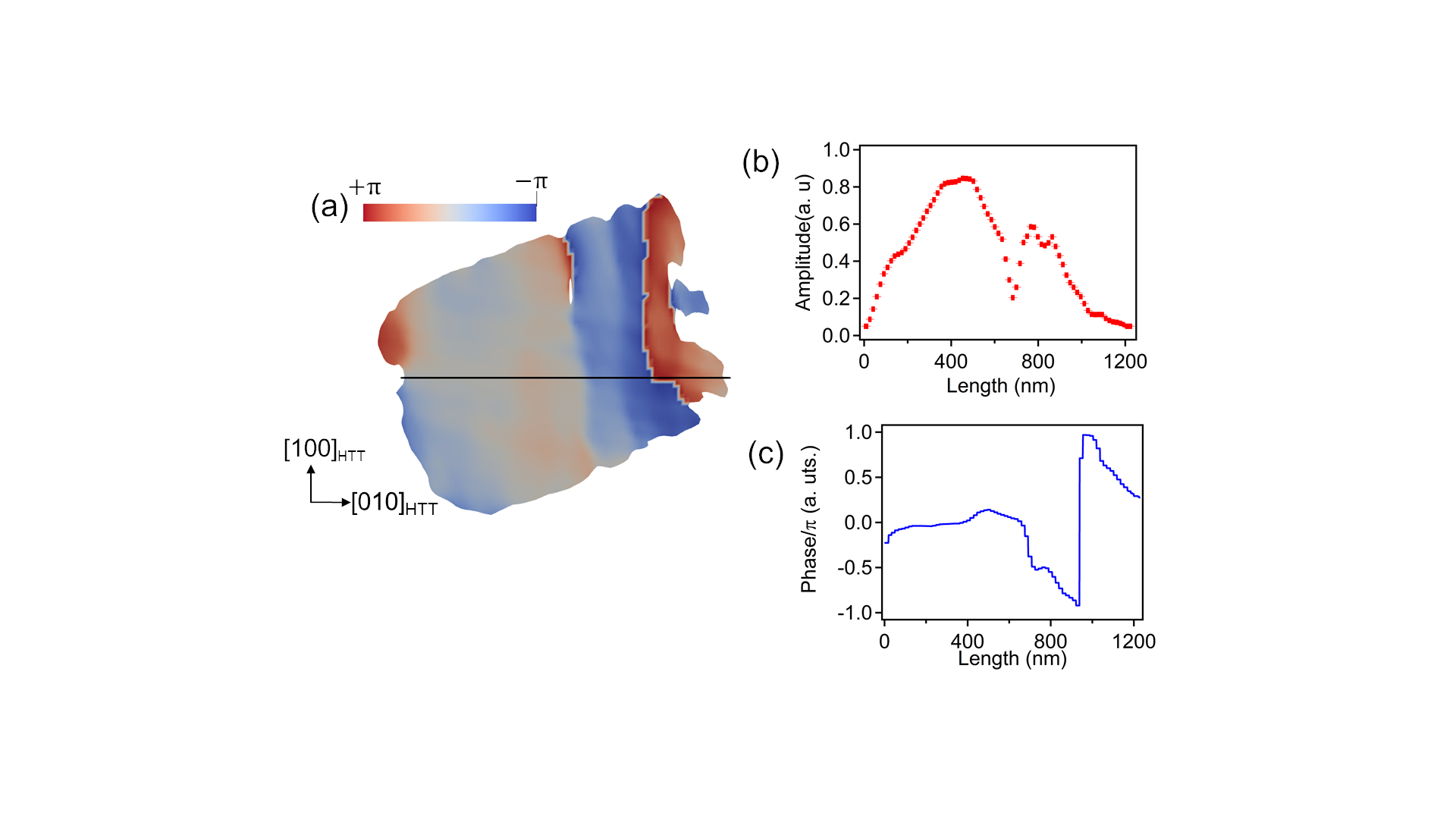}
	\caption{\label{fig:FigS3} Two-dimensional slices of [001] plane of the 258 K data. (a) and (c) are line-profiles of the amplitude (electron density) and normalized phase (projection of the lattice displacement) along the [010]$_{\textrm{HTT}}$ direction respectively. Black line shows the position where line-profile data is taken.}
\end{figure}

\vspace{2cm}
\begin{figure}[htb!]
	\includegraphics[trim={2.5cm 1.5cm 1.5cm 1.5cm},clip,scale=1.6]{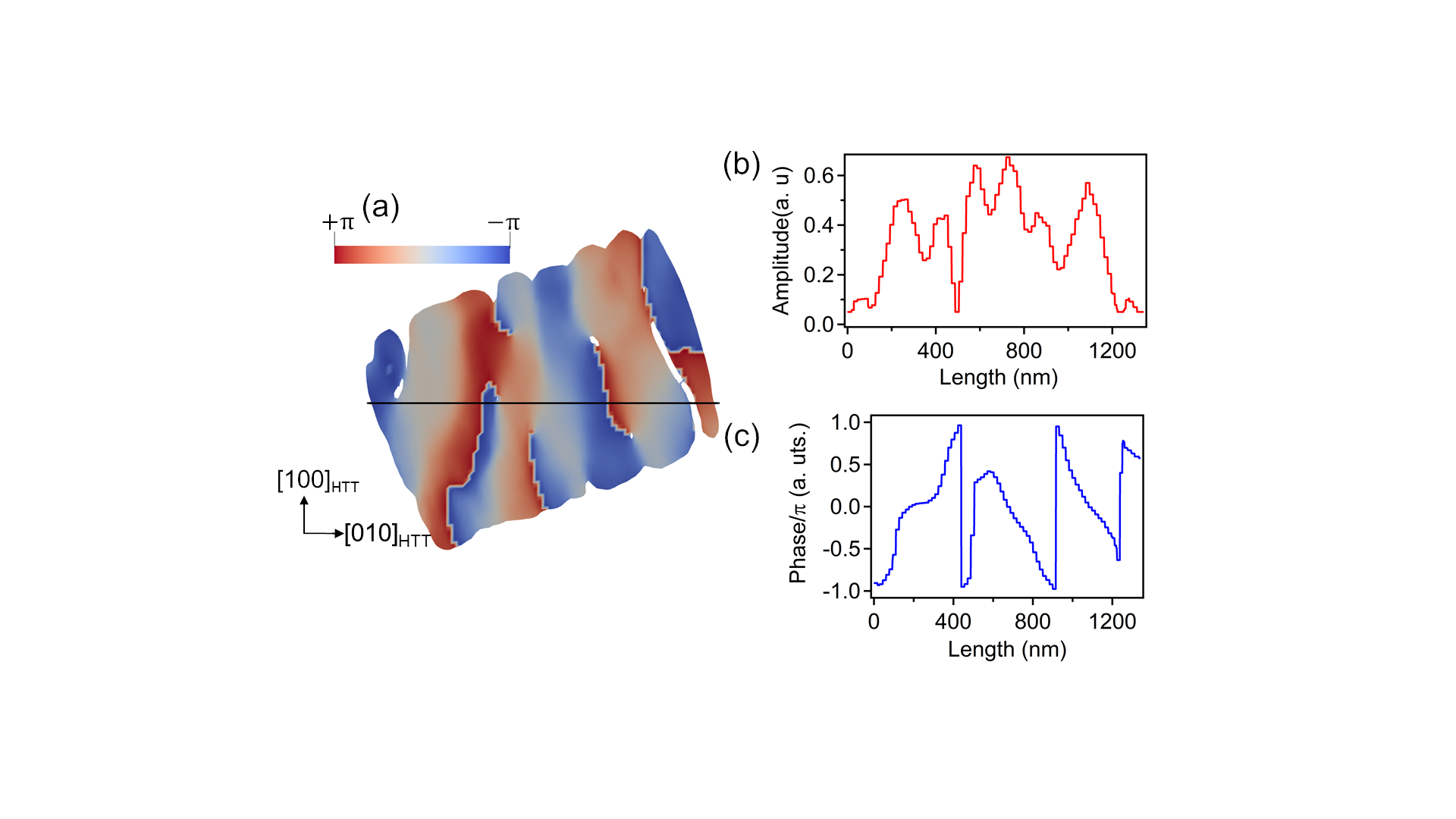}
	\caption{\label{fig:FigS4} Two-dimensional slices of [001] plane of the 228 K data. (a) and (c) are line-profiles of the amplitude (electron density) and normalized phase (projection of the lattice displacement) along the [010]$_{\textrm{HTT}}$ direction respectively. Black line shows the position where line-profile data is taken.}
\end{figure}

\end{document}